\documentclass[pre,aps,superscriptaddress,showpacs]{revtex4-1}
\usepackage{natbib}
\usepackage{amsmath}
\usepackage{epsfig}
 	
\begin{document}
\title{Calculating work in adiabatic two-level quantum Markovian master equations: a characteristic function method}
\author{Fei Liu}
\email[Email address: ]{feiliu@buaa.edu.cn} 
\affiliation{School of Physics and Nuclear Energy Engineering, Beihang University, Beijing 100191, China}
\date{\today}

\begin{abstract}
{We present a characteristic function method to calculate the probability density functions of the inclusive work in the adiabatic two-level quantum Markovian master equations. These systems are steered by some slowly varying parameters and the dissipations may depend on time. Our theory is based on the interpretation of the quantum jump for the master equations. In addition to the calculation, we also find that the fluctuation properties of the work can be described by the symmetry of the characteristic functions, which is exactly the same as the case of the isolated systems. A periodically driven two-level model is used to show the method.}
\end{abstract}
\maketitle

\section{Introduction}
In the past decade, extending classical work equalities\cite{Bochkov1977,Jarzynski1997,Crooks1999,Crooks2000} into the  nonequilibrium quantum regime has attracted intensive interest~\cite{Bochkov1977,Kurchan2000,Tasaki2000,
Yukawa2000,Piechocinska2000,Allahverdyan2005,Talkner2007,Andrieux2008,Talkner2009,Esposito2009,Campisi2011,Campisi2011a,Liu2012,Rastegin2014,DeRoeck2007,Subasi2012,Horowitz2013,Mukamel2003,DeRoeck2004,Esposito2006,Crooks2008,
Horowitz2012,Chetrite2012,Hekking2013,Leggio2013,Albash2013,Liu2014}. With the growing consensus about the definitions of work and their equalities in the isolated quantum systems~\cite{Campisi2011}, recently, some attentions were devoted to the quantum Markovian master equations (QMMEs)~\cite{Mukamel2003,DeRoeck2004,Esposito2006,Crooks2008,Horowitz2012,Chetrite2012,Hekking2013,Liu2014,Leggio2013,Albash2013,Horowitz2013}. 
Among them, the notion of quantum jump~\cite{Carmichael1993,Plenio1998,Breuer2002,Wiseman2010} in the quantum optics literature was introduced. As one of statistical interpretations of the master equations having the Lindblad form~\cite{Davies1974,Lindblad1976,Gorini1976}, the quantum jump not only provides the physically reasonable definitions about work for these quantum systems, but also makes the quantum extensions of the work equalities straightforward. For instance, combining this notion with the two energy measurements scheme~\cite{Kurchan2000,Campisi2011a}, Horowitz~\cite{Horowitz2012} proved a quantum Jarzynski equality (QJE) for a specific type of master equations. These equations were assumed to have instantaneous thermal equilibrium solutions. With a similar idea Hekking and Pekola~\cite{Hekking2013} and we~\cite{Liu2014} presented a quantum Bochkov-Kuzovlev equality (BKE)~\cite{Bochkov1977} for another type of master  equations. Different from those in Ref.~\cite{Horowitz2012}, the systems of the latter are driven by weak external fields and their dissipations are time-independent.  

Although these achievements are significant, we notice that most of them focused on the formal derivations about the work equalities in the various QMMEs; few~\cite{Hekking2013,Liu2014} investigated the calculations of the probability density functions (pdfs) of work. In our opinion, this kind of efforts is essential since the pdf of work is fundamental in the thermodynamics of the finite quantum systems~\cite{Allahverdyan2005}. The work equality is only one of the characters of work under specific conditions~\cite{Jarzynski2011,Campisi2011}. A direct method of calculating work is the simulation~\cite{Hekking2013}. By repeatedly generating the quantum jumps~\cite{Plenio1998,Breuer2002,Wiseman2010}, one may readily construct the statistic histograms of work. However, it is inconvenient for theoretical investigations. For instance, the simulation does not provide us with  a relation between the moment of work and the master equation. Additionally, it also bears the errors of statistical sampling. Very recently, in a specific type of master equations an alternative method was developed by us~\cite{Liu2014}. It is based on solving the characteristic function (CF) of the exclusive work~\cite{Jarzynski2007}. This method not only presents the closed expressions of the moments of the work, but also is simple in the numerical realization. Due to these attractive features, in this paper we try to extend the previous CF method to the case of the inclusive work~\cite{Jarzynski2007} in the quantum adiabatic master equations~\cite{Davies1978,Alicki1979,Albash2012,Rousochatzakis2005,Cai2010}. These equations describe the dynamics of the dissipated systems that are adiabatically steered by some external parameters. They were often utilized to model the decoherence effects of the thermal environments in the quantum adiabatic  computation~\cite{Alicki2006,Vega2010,Amin2008,Childs2001,Boixo2013}.

The paper is organized as follows. In Sec.~\ref{notations} we briefly review a generic two-level adiabatic QMME and its quantum jump interpretation. The essential notations are set up. In Sec.~\ref{BackwardQMMEs} we define the backward master equation of the forward equation. In Sec.~\ref{c&qJE} we prove that the QJE in the same forward master equation possesses two different expressions. On the basis of this observation, in Sec.~\ref{characteristicfunction} we present the CF method to calculate the pdfs of the inclusive work. In Sec.~\ref{application} a simple two-level model is used to illustrate our method. Section~\ref{conclusion} concludes this paper. Some key points in the formal derivations are shown in Appendix I and II.

\section{Two-level adiabatic QMME and quantum jump interpretation}
\label{notations}
For simplicity in notations, throughout this paper we employ a generic two-level adiabatic master equation to develop our theory. The most general form of the equation can be found in Ref.~\cite{Albash2012}. Although we will use the Pauli matrices, we do not consider physical spins. In the time interval $(0,t_f)$, the two-level system (TLS) evolves under an adiabatically varying Hamiltonian $H(t)$. Meanwhile, it exchanges energy with a heat bath at the inverse temperature $\beta$. The time-dependence is usually implemented by some external parameters. Here we did not explicitly write them out. We assume the interaction term between the system and the heat bath to be $H_I=A\otimes B$. Under the adiabatic condition, the weak-coupling Markovian approximation, and the secular approximation, the equation of motion of the reduced density matrix $\rho(t)$ for the system is~\cite{Davies1978,Alicki1979,Albash2012,Rousochatzakis2005,Cai2010}:
\begin{eqnarray}
\label{QAMME}
\partial_t\rho(t)&=&{\cal L}_{t}\rho(t)=-\frac{i}{\hbar}[H(t) ,\rho(t)]+D_t[\rho(t)].
\end{eqnarray}
The time-dependent dissipation term is 
\begin{eqnarray}
\label{dissipationterm}
D_t[\rho]&=&\sum_{\alpha=\pm}\gamma_\alpha(\omega_t)\left[A_{\alpha}(t)\rho A^\dag_{\alpha}(t)-\frac{1}{2}\left\{ A^\dag_{\alpha}(t)A_{\alpha}(t),\rho\right \}\right]+\gamma_0\left[A_0(t)\rho A^\dag_0(t)-\frac{1}{2}\left\{A^\dag_0(t)A_0(t),\rho\right\}\right].
\end{eqnarray}
The rates $\gamma_{\pm}(\omega)$ and $\gamma_0$ equal  $\Gamma(\mp\omega)$ and $\Gamma(0)$, respectively, where  $\Gamma(\omega)$=$\int_{-\infty}^{+\infty}d\tau e^{i\omega\tau}\langle B(\tau)B(0)\rangle_{eq}$ and the average is associated with the equilibrium heat bath. The Lindblad operators $A_{\pm}(t)$ and $A_0(t)$ are
\begin{eqnarray}
|\varepsilon_\pm(t)\rangle
\langle \varepsilon_\pm(t) |A |\varepsilon_\mp(t)\rangle
\langle \varepsilon_\mp(t) | ,
\end{eqnarray}
and 
\begin{eqnarray}
\sum_{\alpha=\pm} |\varepsilon_\alpha(t)\rangle
\langle \varepsilon_\alpha(t) |A |\varepsilon_\alpha(t)
\rangle\langle \varepsilon_\alpha(t) |,
\end{eqnarray}
respectively, where $|\varepsilon_\pm (t)\rangle$ are the adiabatic (instantaneous) eigenvectors of $H(t)$ with eigenvalues $\varepsilon_\pm(t)$. These operators have the properties: $A^\dag_{\pm}(t)=A_{\mp}(t)$, $A_0^\dag(t)=A_0(t)$, 
\begin{eqnarray}
\label{eigenoperators}
[H(t),A_\pm(t)]=\pm\hbar\omega_t A_\pm(t),
\end{eqnarray}
and $[H(t),A_0(t)]=0$, where $\hbar\omega_t$$=$$\varepsilon_+(t)$$-$$\varepsilon_-(t)$. 
The crucial assumption on which this paper depends is the instantaneous detailed balance condition, $\gamma_+(\omega_t)=\gamma_-(\omega_t)e^{-\beta \hbar \omega_t}$. In addition, we also specify the correlation function of the heat bath to be an Ohmic spectral density~\cite{Leggett1987}, {\it i.e.}, $\gamma_0=\kappa/\hbar\beta$ and 
$\gamma_-(\omega)=\kappa\omega/(1-e^{-\beta\hbar\omega})$, where $\kappa$ is the coupling strength. The structure of Eq.~(\ref{QAMME}) and the instantaneous detailed balance condition ensure that the TLS always has an instantaneous thermal state 
\begin{eqnarray}
\rho_{eq}(t)=\sum_{\alpha=\pm}\frac{e^{-\beta \varepsilon_\alpha(t)}}{Z(t)}|\varepsilon_\alpha(t)\rangle\langle\varepsilon_\alpha(t)|=\sum_{\alpha=\pm}p^{eq}_\alpha (t)|\varepsilon_\alpha(t)\rangle\langle\varepsilon_\alpha(t)|,
\end{eqnarray}
where $Z(t)={\rm Tr}[e^{-\beta H(t)}]$ is the instantaneous partition function at time $t$. Finally, we specify the initial density matrix to be $\rho_{eq}(0)$ unless otherwise stated. The conditions for the physical validity of Eq.~(\ref{QAMME}) have been rigorously analyzed~\cite{Albash2012,Kamleitner2013}. 

According to the quantum jump theory~\cite{Plenio1998,Breuer2002,Wiseman2010}, the density matrix $\rho(t)$ can be interpreted as a statistical average of the wave function $\psi(t)$. This wave function varies in the Hilbert space of the TLS by alternatively deterministic continuous evolution and stochastic jumps. Its deterministic equation of motion is 
\begin{eqnarray}
\label{determinedwavefunction}
\partial_t\psi(t)&=&-\frac{i}{\hbar}\hat{H}(t)\psi(t)\nonumber\\
&=&-\frac{i}{\hbar}H(t)\psi(t) -\frac{1}{2}\left(\sum_{\alpha=\pm}\gamma_\alpha(\omega_t)
A^\dag_\alpha(t)A_\alpha(t)+\gamma_0A^\dag_0(t)A_0(t)\right)\psi(t).
\end{eqnarray}
Occasionally, the continuous evolution is interrupted by a jump to one of the three states: $A_{\pm}(t)\psi(t)/\| A_{\pm}(t)\psi(t)\|$ and $A_0(t)\psi(t)/\| A_0(t)\psi(t)\|$. We name them $A_{\pm}$- and $A_0$-jumps, respectively.  The probabilities of these jumps are proportional to $\gamma_{\pm}(\omega_t)\|A_{\pm}(t)\psi(t)\|^2$ and $\gamma_0\|A_0(t)\psi(t)\|^2$, respectively. Since the wave function $\psi(t)$ can be always written as $\sum_{\alpha=\pm}c_\alpha (t)|\varepsilon_\alpha(t)\rangle$, after jump the former two states are indeed $|\varepsilon_\pm(t)\rangle$ and their jumping probabilities are proportional to $\gamma_\pm(\omega_t) | c_\mp(t)|^2$. From the energetic point of view, $A_{\pm}$-jumps accompany an absorption and a release of an energy $\hbar\omega_t$ by the system from and to the heat bath, respectively. On the contrary, the $A_0$-jump only induces the changes of the local phases of the wave function. Given the above explanations, the probability of observing a trajectory in the time interval $(0,t)$ which its initial state is $|\psi_0\rangle$, undergoes $N$ jumps at increasing times $t_i$ ($i$$=$$1$,$\cdots$,$N$)  with an order of jumps $(A_{\alpha_1},\cdots,A_{\alpha_N})$ is  
\begin{eqnarray}
\label{trajprob}
&&\prod_{i=N}^1dt_i\prod_{i=N}^1
\gamma_{\alpha_i} \|{\cal L}_N(t,0) |\psi_0\rangle \|^2\nonumber\\
&=&\prod_{i=N}^1dt_i \prod_{i=N}^1
\gamma_{\alpha_i}  \|U(t,t_{N})A_{\alpha_N}(t_N)\cdots
U(t_2,t_1)A_{\alpha_1}(t_1) U(t_1,0) |\psi_0\rangle \|^2.
\end{eqnarray}
Here ${\alpha_i}$ equals $\pm$ or $0$. We did not explicitly write out $\omega_{t_i}$ in the rates $\gamma_{\pm}$. Additionally, the notation $U$ is the {\it non-unitary} time evolution operator of Eq.~(\ref{determinedwavefunction}) in a certain time interval, {\it e.g.}, $U(t_2,t_1)={\cal T}_{-}\exp[-\frac{i}{\hbar}\int_{t_1}^{t_2}d\tau\hat{H}(\tau)]$, where  ${\cal T}_-$ denotes the chronological time-ordering operator. With the probability density of the quantum trajectory and doing a summation over all trajectories, one may calculate the density matrix by the wave-function as  $\rho(t)=E\left[|\psi(t)\rangle\langle \psi(t)|\right]$~\cite{Plenio1998,Breuer2002,Wiseman2010}. 

\section{Backward adiabatic QMMEs}
\label{BackwardQMMEs}
The work equalities are intimately related to the symmetry of the system and its time-reversal~\cite{Jarzynski2011,Crooks1999,Crooks2000,Campisi2011}. As a preliminary of the following discussion, we introduce the time-reversal of the forward Eq.~(\ref{QAMME}), or the backward adiabatic QMME. First we define $\widetilde{H}(s)=\Theta H(t)\Theta^\dag$ as the time-reversed Hamiltonian, where another time parameter $s=t_f-t$, and $\Theta$ is the time-reversal operator. Throughout this paper, we use the notations with {\it tilde} to denote their meanings under the time-reversal. Obviously, the eigenvectors $|\widetilde\varepsilon_\alpha(s)\rangle$ and eigenvalues $\widetilde\varepsilon_\alpha (s)$ of $\widetilde H(s)$ equal $\Theta |\varepsilon_\alpha(t)\rangle$ and $\varepsilon_\alpha(t)$, respectively. Given the interaction Hamiltonian $H_I$ is time-reversible, which we always assume here, we introduce the backward master equation  
\begin{eqnarray}
\label{BQAMME}
\partial_s\widetilde\rho(s)&=&\widetilde{\cal L}_{s}\widetilde\rho(s)=-\frac{i}{\hbar}\left[\widetilde H(s) ,\widetilde \rho(s)\right]+\widetilde D_s\left[\widetilde\rho(s)\right].
\end{eqnarray}
The dissipation term is 
\begin{eqnarray}
\label{bdissipationterm}
\tilde D_s[\widetilde\rho]&=&\sum_{\alpha=\pm}\widetilde \gamma_\alpha(\widetilde\omega_s)\left[\widetilde A_{\alpha}(s)\widetilde \rho \widetilde A^\dag_{\alpha}(s)-\frac{1}{2}\left\{\widetilde A^\dag_{\alpha}(s)\widetilde A_{\alpha}(s),\widetilde\rho\right\}\right]+\widetilde \gamma_0\left[\widetilde A_0(s)\widetilde \rho \widetilde A^\dag_0-\frac{1}{2}\left\{\widetilde A^\dag_0(s)\widetilde A_0(s),\widetilde\rho\right\}\right]. 
\end{eqnarray}
The time-reversed rates and Lindblad operators have simple connections with the original ones: $\widetilde\gamma_\alpha(\widetilde\omega_s)=\gamma_\alpha(\omega_t)$, $\widetilde\gamma_0=\gamma_0$, $\widetilde A_{\pm}(s)=\Theta A_{\pm}(t)\Theta^\dag$ and $\widetilde A_0(s)=\Theta A_0(t)\Theta^\dag$. Compared Eq.~(\ref{BQAMME}) with~(\ref{QAMME}), we see that the former may be obtained from the latter by replacing $t$ by $s$ and adding tildes on all relevant quantities therein. Because the backward equation is still adiabatic, it has the interpretation of the quantum jump as well. For instance, the deterministic evolution equation for Eq.~(\ref{BQAMME}) is,
\begin{eqnarray}
\label{Rdeterminedwavefunction}
\partial_s\widetilde\psi(s)&=&-\frac{i}{\hbar}\hat{\widetilde{H}}(s)\widetilde\psi(s),
\end{eqnarray}
where $\hat{\widetilde{H}}(s)$ is analogous to $\hat{H}(t)$ in Eq.~(\ref{determinedwavefunction}) except that the operators and rates therein are replaced by their time-reversals.
  
\section{Two expressions of QJE}
\label{c&qJE}
In order to construct the CF method about the inclusive work, we first prove the equivalence of two QJEs in the same master equation~(\ref{QAMME}). They were proposed by Horowitz~\cite{Horowitz2012} and Chetrite and Mallick~\cite{Chetrite2012}, respectively. The latter  equality is an abstract ``book-keeping" of a sum of multiple time correlation functions of the operators. Interestingly, the notions of the quantum jump and two energy measurements were not involved. Following our previous convention~\cite{Liu2014}, we name them the $c$- and $q$-$number$ QJEs, respectively. So far, their relation was not clarified. The reader will see that the equivalent demonstration indeed provides us with a shortcut toward an important evolution equation that can  assist the calculation of the CF. 
 
\subsection{$c$-$number$ QJE} 
\begin{figure}
\includegraphics[width=1\columnwidth]{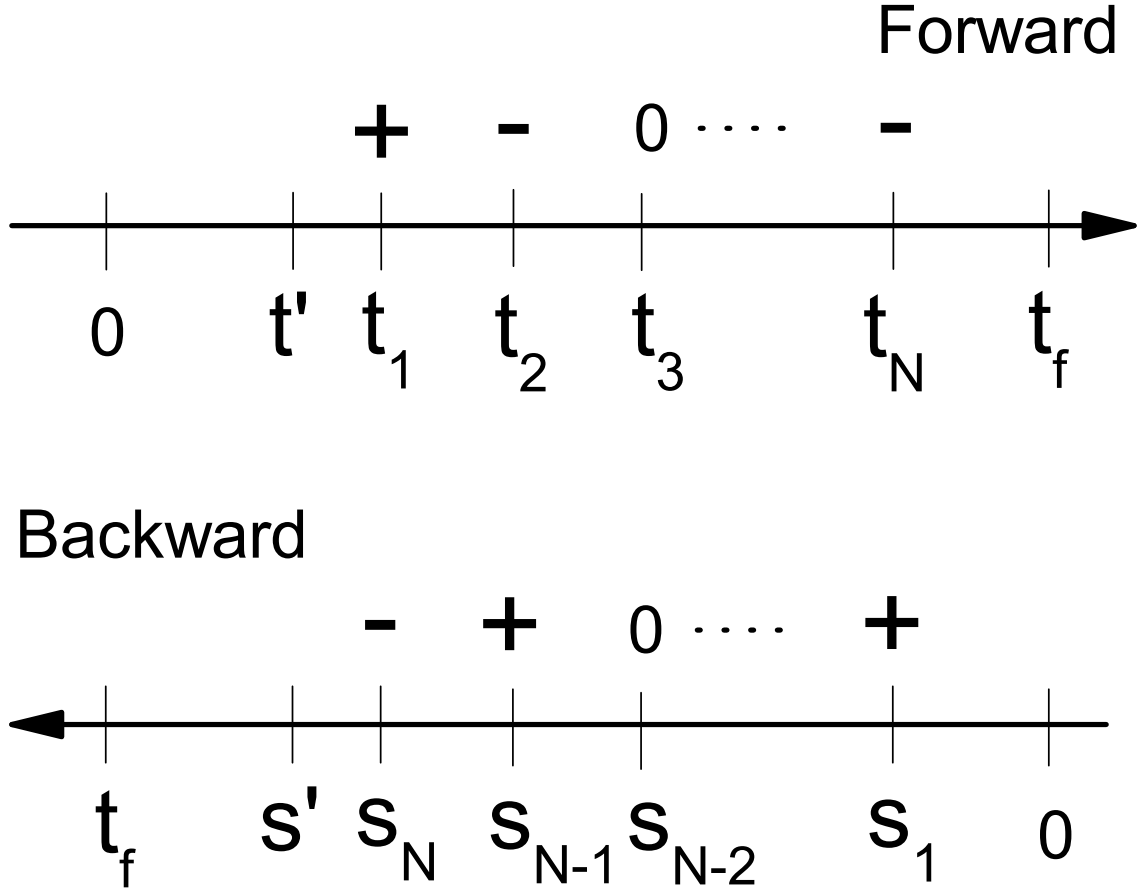}
\caption{A quantum jump trajectory of the forward adiabatic QMME and its time reversal. The arrows indicate the directions of time. The symbols $\pm$ and $0$ represent the $A_{\pm}$- and $A_0$-jumps, respectively. }\label{figure1}
\end{figure}

Let us choose an arbitrary time $t'$ between 0 and $t_f$ and suppose that the wave function at the time is $|\varepsilon_{\alpha}(t')\rangle$. Given a quantum trajectory of Eq.~(\ref{QAMME}) starting with this state. If we record the order of jumps $(A_{\alpha_1},\cdots,A_{\alpha_N})$ at later times $(t_1,\cdots,t_N)$, and measure the energy eigenvector of the TLS at the terminal time $t_f$ to be $|\varepsilon_\delta(t_f)\rangle$, we define the inclusive work done on the system along the trajectory in the time interval $(t',t_f)$ as
\begin{eqnarray}
\label{workdef} W(t')=\varepsilon_\delta(t_f)-\varepsilon_{\alpha}(t')-
\int_{t'}^{t_f}\hbar\omega_{\tau} dN_+(\tau)+\int_{t'}^{t_f}\hbar\omega_{\tau} dN_-(\tau),
\end{eqnarray}
where $dN_\pm(\tau)$ represent the increments of the $A_\pm$-jumps at time $\tau$. Note that $N_0$ the number of the $A_0$-jumps is not involved since they do not contribute any energy changes. Now we are concerned about the following equation,
\begin{eqnarray}
\label{jointprob}
e^{-\beta W(t')}\left[\prod_{i=N}^1dt_i\prod_{i=N}^1
\gamma_{\alpha_i} \|\langle \varepsilon_\delta(t_f)| {\cal L}_N(t_f,t')
|\varepsilon_\alpha(t')\rangle \|^2\right]  p^{eq}_\alpha(t').
\end{eqnarray}
According to Eq.~(\ref{trajprob}), the whole term in the above square brackets is the conditional probability of observing the trajectory. Hence, its product with $p^{eq}_\alpha(t')$ is the joint probability. Equation~(\ref{jointprob}) possesses an intriguing explanation of time reversal~\cite{Crooks2008,Horowitz2012}. We first notice that the terms in the exponential function of the work can be combined into the rates using the instantaneous detailed balance condition. Then we rewrite the equation as
\begin{eqnarray}
\label{TRjointprob}
\frac{Z(t_f)}{Z(t')} \left[\prod_{i=1}^Ndt_i\prod_{i=1}^N\gamma_{\widetilde{\alpha_i}} \|\langle \varepsilon_\alpha(t')|\Theta^\dag \widetilde{\cal L}_N(s',0)\Theta
|\varepsilon_\delta(t_f)\rangle \|^2\right] p^{eq}_\delta(t_f),
\end{eqnarray} 
where $\widetilde{\alpha_i}$ denotes $\mp$ or $0$ if $\alpha_i$ is $\pm$ or 0, respectively. The operator $\widetilde{{\cal L}}_N(s',0)$ with $t'+s'=t_f$ is 
\begin{eqnarray}
\label{reversedwholeevolutionoperator}
[\Theta U^\dag(t_1,t')\Theta^\dag][\Theta A^\dag_{\alpha_1}(t_1)\Theta^\dag]  \cdots [\Theta U^\dag(t_N,t_{N-1})\Theta^\dag][\Theta A^\dag_{\alpha_N}(t_N) \Theta^\dag] [\Theta U^\dag(t_f,t_{N})\Theta^\dag].
\end{eqnarray}
We immediately see that the term in the second square brackets is just $\widetilde A_{\widetilde{\alpha_1}}(s_N)$ of the backward Eq.~(\ref{BQAMME}). Here we define $s_j+t_i$$=$$t_f$ and $i+j$$=$$N+1$. Note that the $\widetilde A_{\widetilde{\alpha_1}}$- and $A_{\alpha_1}$-jumps are opposite unless $\alpha_1=0$. Moreover, we may check that the term in the first square brackets is the non-unitary time evolution operator $\widetilde{U}(s',s_{N})$ of Eq.~(\ref{Rdeterminedwavefunction}) in the time interval $(s_N,s')$. For the remaining terms in Eq.~(\ref{reversedwholeevolutionoperator}) these two observations are true as well. Noting $\widetilde\gamma_{\widetilde\alpha_i}(\widetilde\omega_{s_i})=\gamma_{\widetilde\alpha_i}(\omega_{t_j})$, we finally find that the whole term in the square brackets of Eq.~(\ref{TRjointprob}) is nothing but the conditional probability of a quantum trajectory for the backward master equation~(\ref{BQAMME}):  its state at time 0 is $\Theta|\varepsilon_\delta(t_f)\rangle$, the order of jumps is $(\widetilde A_{\widetilde{\alpha_N}},\cdots,\widetilde A_{\widetilde{\alpha_1}})$ at times $(s_1,\cdots,s_N)$, and the energy eigenvector measured at the final time $s'$ is $\Theta|\varepsilon_\alpha(t')\rangle$. Fig.~(\ref{figure1}) is a schematic diagram of two time-revered quantum trajectories. 

Now we do a summation of  Eq.~(\ref{jointprob}) over all quantum trajectories that start with the same $|\varepsilon_{\alpha}(t')\rangle$ and end at all the energy eigenvectors. Using Eq.~(\ref{TRjointprob}), we establish an important equation 
\begin{eqnarray}
\label{forwardbackwardrelation}
&&E_\alpha[e^{-\beta W(t')}]p^{eq}_{\alpha}(t')=\frac{Z(t_f)}{Z(t')}\langle \varepsilon_{\alpha}(t')| \Theta^{\dag}
\widetilde{\rho}(s') \Theta|\varepsilon_{\alpha}(t')\rangle. 
\end{eqnarray}
We used $E_{\alpha}$ to denote that all trajectories start with the same quantum state. The reduced density matrix $\widetilde\rho(s')$ is the solution of Eq.~(\ref{BQAMME}) at time $s'$. Particularly, its initial condition $\widetilde\rho(0)$ has been specified at the thermal state $\Theta\rho_{eq}(t_f)\Theta^\dag$. If we further sum Eq.~(\ref{forwardbackwardrelation}) over the index $\alpha$ and choose $t'=0$, the $c$-$number$ QJE in the two-level adiabatic master equation~(\ref{QAMME}) is obtained: 
\begin{eqnarray}\label{cQJE}
E[e^{-\beta W(0)}]=e^{-\beta\Delta G},
\end{eqnarray}
where $\beta\Delta G=\ln{Z(t_f)}-\ln{Z(0)}$. The reader is reminded that the initial condition of the forward Eq.~(\ref{QAMME}) must be the thermal state, which explains why we set up this condition at the beginning.

\subsection{$q$-$number$ QJE}
Equation~(\ref{forwardbackwardrelation}) implies that it may arise from an almost trivial operator identity: 
\begin{eqnarray}
\label{Roperator}
R(t',t_f)\rho_{eq}(t')=\Theta^\dag \widetilde{\rho}(s')\Theta.
\end{eqnarray}
Note that its validity has nothing to do with the quantum jump provided the well-defined $\widetilde\rho(s')$ and $\rho_{eq}(t')$. Writing Eq.~(\ref{Roperator}) in the energy representation and comparing it with Eq.~(\ref{forwardbackwardrelation}), we have
\begin{eqnarray}\label{connectionR&average}
\langle\varepsilon_{\alpha}(t')|R(t',t_f)|\varepsilon_{\alpha}(t')\rangle=\frac{Z(t')}{Z(t_f)}E_\alpha [e^{-\beta W(t')}].
\end{eqnarray}
Obviously, the operator $R(t',t_f)$ possesses all characters that the $E_\alpha$-term owns. In the following we pay our attention on the general properties of $R(t',t_f)$ and temporarily set aside the quantum jump. Substituting Eq.~(\ref{Roperator}) into Eq.~(\ref{BQAMME}), we are able to obtain an evolution equation about $R(t',t_f)$ with respect to $t'$
\begin{eqnarray}
\label{evolutioneqofR}
\partial_{t'}R(t',t_f)=-{\cal L}^{\star}_{t'} R(t',t_f)-R(t',t_f)\partial_{t'}\rho_{eq}(t') \rho^{-1}_{eq}(t') ,
\end{eqnarray}
where the adjoint superoperator of ${\cal L}_{t'}$ is 
\begin{eqnarray}
\label{adjointoperator}
{\cal L}_{t'}^\star {O}=\frac{i}{\hbar}\left[H(t') ,O\right]&+&\sum_{\alpha=\pm}\gamma_\alpha(\omega_{t'})
\left[A^\dag_{\alpha}(t'){O} A_{\alpha}(t')-\frac{1}{2}\left\{A^\dag_{\alpha}(t')A_{\alpha}(t'),{O}\right\}\right]\nonumber\\
&+&\gamma_0\left [A^\dag_0(t'){O} A_0(t')-\frac{1}{2}\left\{A^\dag_0(t')A_0(t'),{O}\right\}\right].
\end{eqnarray}
To arrive at Eq.~(\ref{evolutioneqofR}), we applied the instantaneous detailed balance condition again. Here we must emphasize that this result does not matter with the initial conditions of the forward and backward master equations. In addition, Eq~(\ref{evolutioneqofR}) is a terminal value problem, {\it i.e.}, $R(t_f,t_f)=I$ the identity operator. Introducing the adjoint propagator $G^\star(t_1,t_2)$$=$${\cal T}_{+}\exp [\int_{t_1}^{t_2}d\tau {\cal L}^\star_{\tau}$]~\cite{Breuer2002}
 $(t_1$$<$$t_2)$, where ${\cal T}_{+}$ denotes the antichronological time-ordering operator, we may have a formal solution of  Eq.~(\ref{evolutioneqofR}) written by the celebrated Dyson series~\cite{Chetrite2012,Liu2012a,Liu2014}. Choosing $t'=0$ and taking traces on two sides of Eq.~(\ref{Roperator}),  we obtain the $q$-$number$ QJE~\cite{Chetrite2012,Liu2012a} for the same master equation~(\ref{QAMME})
\begin{eqnarray}
\label{qQJE}\left\langle {\cal T}_{+}\exp\left[\int_0^{t_f} d\tau{\cal W}(\tau)\right]\right\rangle=e^{-\beta\Delta G}. 
\end{eqnarray}
Here we defined an operator ${\cal W}(\tau)=\partial_{\tau}e^{-\beta H(\tau)}e^{\beta H(\tau)}$. It is worthy to point out that the ``average" $\langle$ $\rangle$ above is only a shorthand notation~\cite{Chetrite2012}. Indeed, its explicit expression is a sum of infinite terms of multiple times correlation functions of the operators~\cite{Breuer2002}:   
\begin{eqnarray}
&&1+\left\langle \int_0^{t_f}dt_1{\cal W}(t_1)\right\rangle+\left\langle \int_0^{t_f}dt_1\int_{t_1}^{t_f}dt_2{\cal W}(t_2){\cal W}(t_1)\right\rangle +\cdots\nonumber\\
=&&1+\int_0^{t_f}dt_1{\rm Tr}\left[{\cal W}(t_1)G(t_1,0)\rho_{eq}(0)\right]+\int_0^{t_f}dt_1\int_{t_1}^{t_f}dt_2{\rm Tr}\left[{\cal W}(t_2)G(t_2,t_1){\cal W}(t_1)G(t_1,0)\rho_{eq}(0)\right]+\cdots,
\end{eqnarray}
where $G(t_2,t_1)={\cal T}_{-}\exp[ \int_{t_1}^{t_2} d\tau{\cal L}_{\tau}]$ is the propagator of Eq.~(\ref{QAMME}). Note that these propagators are superoperators: they act on all terms on their right-hand side.

Equation~(\ref{connectionR&average}) ensures the equivalence of the $c$- and $q$-$number$ QJEs. An alternative proof is to expand the exponential functions in   Eqs.~(\ref{cQJE}) and~(\ref{qQJE}) as a series of the inverse temperature $\beta$ and to check whether their coefficients equal. To implement this scheme, one has to firstly know the correlation functions of the quantum jumps among different times. Fortunately, they have been given previously~\cite{Wiseman2010}. Here we list the final results of the first two coefficients that are indeed the first two moments of the inclusive work: 
\begin{eqnarray}
\label{workexpansion1stmoment} 
E[W]&=&
\int_0^{t_f} dt_1 \left\langle\partial_{t_1}H(t_1)\right\rangle ,  \\
E[W^2]&=&2\int_0^{t_f}dt_1 \int_{t_1}^{t_f}  dt_2
\left\langle\partial_{t_2}H(t_2)\partial_{t_1}H(t_1)\right\rangle +\int_0^{t_f}dt_1 \left\langle[H(t_1),\partial_{t_1}
H(t_1)]\right\rangle.
\label{workexpansion2ndmoment}
\end{eqnarray}
Note that the second term in the second equation is a pure quantum effect. We leave their derivations in the Appendix~I. 

\section{Characteristic function of inclusive work}
\label{characteristicfunction}
In the preceding discussion, we clearly see that the pdf of the inclusive work~(\ref{workdef}) can be constructed using the quantum jump simulation~\cite{Breuer2002,Wiseman2010}. On the contrary, we do not gain such an impression in the case of the $q$-$number$ equality. For the latter, a possible way is to calculate all the moments of the work using the equations like Eqs.~(\ref{workexpansion1stmoment}) and (\ref{workexpansion2ndmoment}) and then to convert them into the pdf. However, for the higher moments, in addition that the higher-dimensional integrations are involved, their connections with the multiple time correlation function will become dramatically complicated. Hence, this method is almost infeasible in practice. 

To bypass this difficulty, we introduce the CF of the pdf, 
\begin{eqnarray}
\label{characterfun} \Phi(\mu)=E\left[e^{i\mu W}\right],
\end{eqnarray}
where $\mu$ is real number. After solving the CF, the pdf  is obtained by performing an inverse Fourier transform of $\Phi(\mu)$. At first glance,  Eq.~(\ref{characterfun}) does not show apparent advantages. However, the CF may be regarded as the left-hand side of the QJE~(\ref{cQJE}) except that $\beta$ therein is replaced by an {\it imaginary} inverse temperature $-i\mu$. Inspired by Eq.~(\ref{connectionR&average}), we want to find an operator $K(t',t_f;\mu)$ analogous to $R(t',t_f)$ by which the CF~(\ref{characterfun}) is calculated as 
\begin{eqnarray}
\label{calculatingCF}
\Phi(\mu)={\rm Tr}\left[K(0,t_f;\mu)\rho_{eq}(0)\right].
\end{eqnarray}
It is not difficult to see that the operator indeed exists and satisfies an evolution equation  
\begin{eqnarray}
\label{evolutioneqCF}
\partial_{t'}K(t',t_f;\mu)=-{\cal L}^{\star}_{t'} K(t',t_f;\mu)- 
K(t',t_f;\mu)\partial_{t'}e^{i\mu H(t')}e^{-i\mu H(t')}, 
\end{eqnarray}
and the terminal condition is $K(t_f,t_f;\mu)=I$. This is the central result of this paper.

At this stage we have achieved the goal of calculating the pdf of the work by solving Eq.~(\ref{evolutioneqCF}) rather than simulating the quantum trajectories. In practice, however, it is inconvenient to compute the exponential functions of the Hamiltonian operator; see the last term in the above equation. In addition, this is a terminal value problem rather than the conventional initial value problem. These two undesirable features may be remedied by introducing another ``better" operator 
\begin{eqnarray}
\widetilde{\cal K}(s';\mu)=\Theta K(t',t_f;\mu)e^{i\mu H(t')}\Theta^\dag.
\end{eqnarray}
After a simple algebra we have 
\begin{eqnarray}
\label{Modifiedevolutioneqcharactersiticfunc}
\partial_{s'}\widetilde{\cal K}(s';\mu)=\widetilde{\breve{\cal L}}_{s'}(\mu) \widetilde{\cal K}(s';\mu),
\end{eqnarray}
and the initial condition $\widetilde{\cal K}(0;\mu)$ equals $e^{-i\mu {\widetilde H}(0)}$. The superoperator of the right-hand side of Eq.~(\ref{Modifiedevolutioneqcharactersiticfunc}) is 
\begin{eqnarray}
\label{CFnewevolutioneq}
\widetilde{\breve{\cal L}}_{s}(\mu){O}=-\frac{i}{\hbar}\left[\widetilde H(s) ,O\right]&+&\sum_{\alpha=\pm}\widetilde \gamma_\alpha(\widetilde\omega_s)\left[e^{\alpha i \mu\hbar\tilde{\omega}_s}\widetilde A^\dag_{\alpha}(s)O\widetilde A_{\alpha}(s)-\frac{1}{2}\left\{\widetilde A^\dag_{\alpha}(s)\widetilde A_{\alpha}(s),O\right\}\right]\nonumber\\
&+&\widetilde \gamma_0\left[\widetilde A^\dag_0(s){O}\tilde A_0(s)-\frac{1}{2}\left\{\widetilde A^\dag_0(s)\widetilde A_0(s),O\right\}\right].
\end{eqnarray}
Accordingly, Eq.~(\ref{calculatingCF}) is slightly modified as 
\begin{eqnarray} 
\label{characteristicfun2ndmethod}
\Phi(\mu)&=&\frac{1}{Z(0)}{\rm Tr}\left[\Theta^\dag \widetilde{\cal K}(t_f;\mu)\Theta e^{i(-\mu+i\beta) H(0)}\right]\nonumber\\
&=&\frac{1}{Z(0)}{\rm Tr}\left[\widetilde{\cal K}(t_f;-\mu)e^{i(-\mu+i\beta) H(0)}\right].
\end{eqnarray}
The second equation is due to ${\rm Tr}[\Theta^\dag O \Theta]={\rm Tr}[O^\dag]$. We see that Eq.~(\ref{CFnewevolutioneq}) is very close to Eq.~(\ref{BQAMME}). Indeed, if we replace all $\mu$ therein by $-i\beta$, the former will reduce into the latter. 

Besides the calculation, Eq.~(\ref{characteristicfun2ndmethod}) is also useful in discussing the symmetry of the pdfs of the inclusive work. On the basis of the quantum jump theory, Horowitz has argued that the Crooks equality~\cite{Crooks2000,Crooks1999} was held in a specific type of master equations~\cite{Horowitz2012}. The equality is about the pdfs of the work for the forward and backward QMMEs. To the end, we first denote the CF for Eq.~(\ref{BQAMME}) to be ${\widetilde \Phi}(\mu)$. We will show that, if the Hamiltonian is time-reversible at arbitrary time, {\it i.e.}, $\Theta H(t)\Theta^\dag=H(t)$, these two CFs satisfy an important symmetry
 \begin{eqnarray}
\label{symmetryCF}
Z(0)\Phi(u)=Z(t_f){\widetilde\Phi}(\nu), 
\end{eqnarray}
where $\nu=i\beta-u$. If one transforms it back into the pdfs, the Crooks equality will be  recovered~\cite{Campisi2011}. We notice that Eq.~(\ref{symmetryCF}) is exactly the same as that in the isolated quantum systems~\cite{Campisi2011}. Because of the duality of the forward and backward equations, for $\widetilde\Phi(\mu)$ we may follow the previous argument to introduce an operator ${\cal K}(t';\mu)$ and require 
\begin{eqnarray} 
\label{characteristicfun2ndmethodReversed}
\widetilde\Phi(\mu) =\frac{1}{Z(t_f)}{\rm Tr}\left[{\cal K}(t_f;-\mu)e^{i(-\mu+i\beta) {\widetilde H}(0)}\right].
\end{eqnarray}
Obviously, the operator ${\cal K}(t';\mu)$ satisfies an evolution equation analogous to Eq.~(\ref{Modifiedevolutioneqcharactersiticfunc}) except that all tildes therein are erased and $s'$ is replaced by $t'$. The symmetry~(\ref{symmetryCF}) is essentially attributed to the relation
\begin{eqnarray}
\label{timereversalpropagators}
\widetilde {\breve{G}}(s,0;\mu)(O)=\Theta {\breve G}^\star(t,t_f;-\nu)(\Theta^\dag O\Theta)\Theta^\dag,
\end{eqnarray}
where $\widetilde {\breve{G}}$ on the left-hand side is the propagator of Eq.~~(\ref{Modifiedevolutioneqcharactersiticfunc}), and ${\breve G}^\star$ on another side is the adjoint propagator of the evolution equation of ${\cal K}(t';\mu)$. Appendix II presents the further details about this  relation. With these notations proving Eq.~(\ref{symmetryCF}) is straightforward:
\begin{eqnarray}
\Phi(\mu)Z(0)&=&{\rm Tr}\left[
\Theta^\dag {\widetilde{\breve G}}(t_f,0;\mu)(\tilde{\cal K}(0;\mu))\Theta
e^{ivH(0)}\right]\nonumber\\
&=&{\rm Tr}\left[{\breve G}^\star(0,t_f;-\nu)(\Theta^\dag{\tilde{\cal K}(0;\mu)}\Theta) e^{i\nu H(0)}\right]\nonumber\\
&=&{\rm Tr}\left[{\breve G}(t_f,0;-\nu)(e^{i\nu H(0)}) e^{i\mu H(t_f)}\right]\nonumber\\
&=&\widetilde{\Phi}(\nu)Z(t_f).
\end{eqnarray}
Note that the last step has used the time-reversible property of the Hamiltonian.

Before closing the theoretical part of this paper, we want to make several comments. The first is the effect of the initial density matrix. So far, we always assumed the initial reduced density matrix $\rho(0)$ to be the thermal state $\rho_{eq}(0)$. However, the inclusive work~(\ref{workdef}) and the characteristic function~(\ref{characterfun}) are always well-defined provided that the initial density matrix is {\it diagonal} in the energy representation, namely, $[\rho(0),H(0)]=0$. Under this circumstance, the calculation of the CF using the evolution Eqs.~(\ref{evolutioneqCF})  or~(\ref{Modifiedevolutioneqcharactersiticfunc}) is still available. One may see this point more clearly in term of the proof of Eqs.~(\ref{workexpansion1stmoment}) and~(\ref{workexpansion2ndmoment}). The second is the relation between the current results and those in the isolated quantum Hamiltonian systems. Obviously, the former reduces into the latter~\cite{Campisi2011} if we impose the interaction Hamiltonian $H_I$ vanishing. Then, all the dissipation terms such as those in Eqs.~(\ref{QAMME}),~(\ref{BQAMME}), and~(\ref{CFnewevolutioneq}) will be absent. Meanwhile, the action of the propagator $G(t_2,t_1)$ on an operator $O$ is simplified into $U(t_2)U^{\dag}(t_1)OU(t_1)U^{\dag}(t_2)$, where $U(t)$ is now the {\it unitary} time evolution operator of $H(t)$. Moreover, it is well worth emphasizing that for the isolated case, one can remove the restriction of the adiabatic evolution of  Hamiltonian that is essential for the physical validity of Eq.~(\ref{QAMME}). We do not pursue the further details here~\cite{Liu2012,Liu2014b}. Finally, Talkner {\it et al.}~\cite{Talkner2009} have used an another characteristic function method to prove the validity of the JE and Crooks' equality for very general open quantum systems. Except for the weak-coupling approximation, the dynamics of the system therein is not required to be Markovian and the external parameters may vary arbitrarily. There are two key ingredients in their method. One is the unitary evolution of the composition of the system and the heat bath. The other is the simultaneous measurements of the energies of the system and the heat bath at the beginning and the end of the process. The theory is fully microscopic while our starting point is the effective dynamics of the reduced system. The advantage of the latter is that it is closer to the real situation in laboratories. Although in principle the results of  Talkner {\it et al.}~\cite{Talkner2009} shall cover what we obtained here if the additional requirements are imposed, we do not think that establishing this connection would be simple from technical perspective; see the analogous efforts in Refs.~\cite{DeRoeck2007,Esposito2009}.

\section{Example}
\label{application}
\begin{figure}
\includegraphics[width=1.\columnwidth]{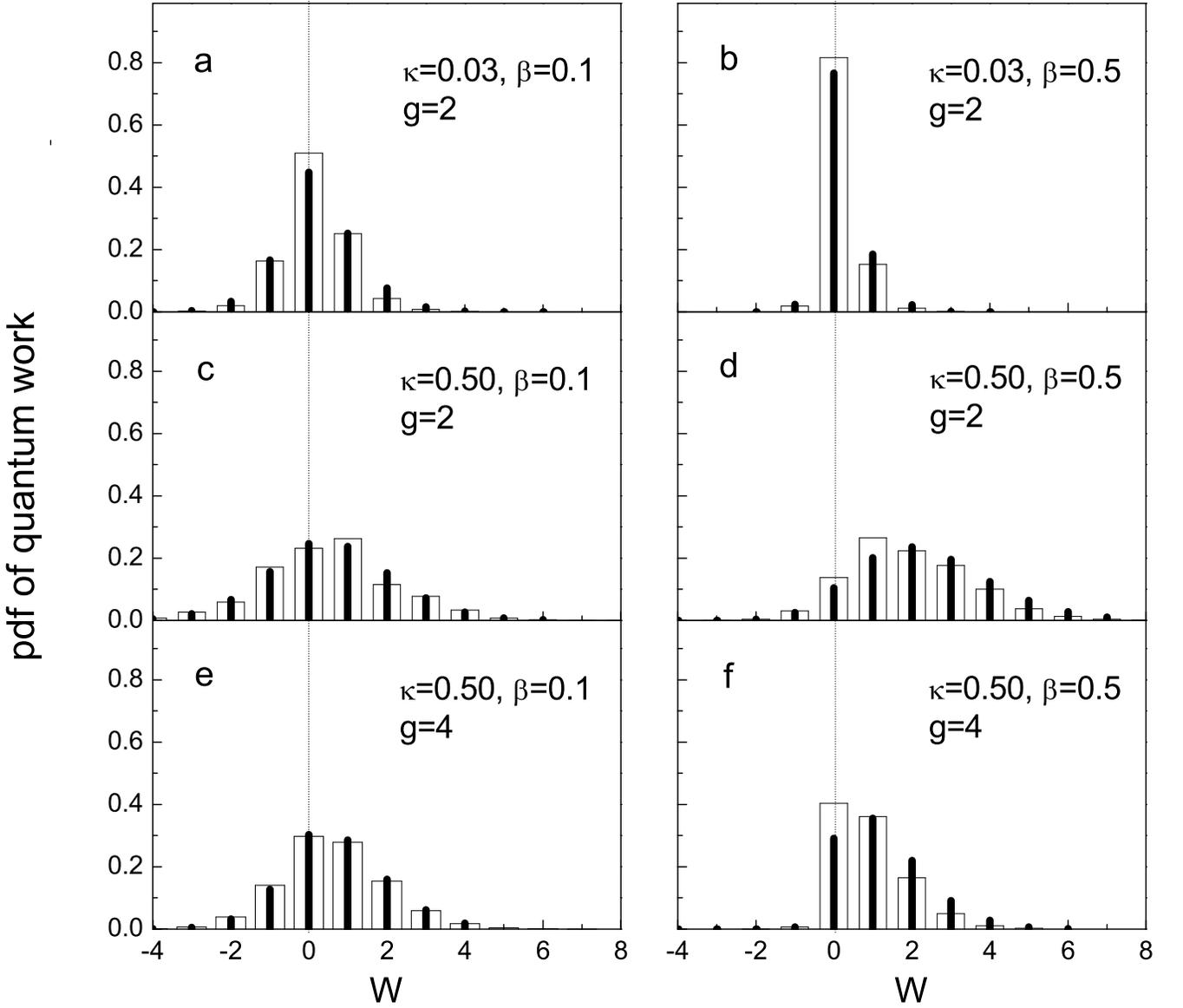}
\caption{The pdfs of the inclusive work for the TLS~(\ref{ATLS}). The bars are calculated by simulating the quantum trajectories, while the solid bold lines are obtained by the CF method. The unit of the work $W$ is $\omega$. In these panels the thin dash lines at zero positions guide for eyes.}
\label{figure2}
\end{figure}

In this section, we will illustrate the CF method by calculating the pdfs of the inclusive work in a simple TLS model. Its Hamiltonian is 
\begin{eqnarray}
\label{ATLS}
H(t)=\frac{1}{2}\hbar\omega_0\sigma_z +g\left(\sigma_+e^{-i\Omega t}+\sigma_-e^{i\Omega t}\right),
\end{eqnarray}
and the operator $A$ is $\sigma_x$. The time-dependent term may be from the rotating wave approximation of the interaction of the TLS with a driving harmonic filed~\cite{Breuer2002}. We simply call $g$ the field strength. We emphasize again that the TLS is not a physical spin. As a result $\sigma_x$ is time-reversible. For the sake of simplicity, we let $\hbar=1$, $k_B=1$, and $\omega_0=1$. The adiabatic eigenvectors and eigenvalues of the Hamiltonian are 
\begin{eqnarray}
\label{eigenvectorU}
|+t\rangle&=&\cos\frac{\theta}{2}e^{-i\Omega t/2}|+\rangle+\sin\frac{\theta}{2}e^{i\Omega t/2}|-\rangle,\\
\label{eigenvectorD}
|-t\rangle&=&-\sin\frac{\theta}{2}e^{-i\Omega t/2}|+\rangle+\cos\frac{\theta}{2}e^{i\Omega t/2}|-\rangle,
\end{eqnarray}
and $\varepsilon_{\pm}=\omega/2$ with a time-independent $\omega$$=$$\sqrt{1+4g^2}$, respectively. Here $|\pm\rangle$ are the two bases of $\sigma_z$ and  $\cos\theta=1/\omega$. The Lindblad operators are 
\begin{eqnarray}
A_+(t)&=&
\left(\cos^2\frac{\theta}{2}e^{i\Omega t}-\sin^2
\frac{\theta}{2} e^{-i\Omega t}\right)|+t\rangle \langle -t|,\\
A_0(t)&=&\sin\theta\cos\Omega t\left(|+t\rangle \langle +t|-|-t\rangle \langle -t|\right).
\end{eqnarray}
The adiabatic condition is very simple: $g\Omega\ll1+4g^2$. Fig.~(\ref{figure2}) shows the pdfs of the inclusive work at different inverse temperatures  $\beta$, the coupling strength $\kappa$, and the field strength $g$. We chose $\Omega=0.99$ and $t_f=20\pi/\Omega$  or 10 cycles. These data are obtained by simulating the quantum jumps and numerically solving the CF under the assistance of the evolution equation~(\ref{Modifiedevolutioneqcharactersiticfunc}), respectively. We see that their agreements are indeed excellent.

These pdfs in Fig.~(\ref{figure2}) can be qualitatively understood from the point of view of the quantum jump. For the TLS the possible values of the change of the system's energy are $\pm\omega$ and 0. If the system is completely isolated, under the adiabatic condition the unique value of the work with nonzero probability is at zero . Let us see the cases in the right column of the figure. Because of the lower temperature (larger $\beta$), we may think of that the wave function $\psi$ of Eq.~(\ref{determinedwavefunction}) always starts with the eigenvector $|-0\rangle$ [Eq.~(\ref{eigenvectorD}) at time 0]. If the system interacts with the heat bath very weakly, {\it e.g.}, $\kappa=0.03$, we expect that the probability of zero work still dominates but there are jumps happing in few quantum trajectories. If in a trajectory a rare jump indeed occurs at some time, it is very possible a $A_-$-jump and an energy $\omega$ is released into the heat bath. The reason is that the rate $\gamma_-$ is far larger than the opposite rate $\gamma_+$ at the lower temperature. The work of these trajectories is $+\omega$. Fig.~(\ref{figure2})b shows this scenario. On the other hand, if we increase $\kappa$ but keep the same temperature, the absolute values of these rates increase while their ratio is still unchanged. In this situation, not only does the population of trajectories with jumps become larger, but the frequency of the jumps in a same trajectory increases. Accordingly, the probability of zero work shall considerably decrease while the probabilities of the work with larger positive values present. This is what we see in Fig.~(\ref{figure2})d. The above arguments also imply that, if we prolong the time $t_f$, we shall see the movements of these pdfs toward the right-hand side. We indeed observe this trend in calculations (data not shown here). Finally, in Fig.~(\ref{figure2})f we see that the probabilities of the negative work almost vanish. This is due to the fact that the larger field strength $g$ leads into the negligible $\gamma_+$. Hence, the $A_+$-jumps that are responsible for the negative work are strongly inhibited. For the cases in the left column of the figure, analogous analyses can be performed. Because of the higher temperature (smaller $\beta$), however, two additional factors must be taken into account. One is that the contribution of the initial state at the eigenvector $|+0\rangle$ becomes significant now. Another is that the two rates $\gamma_+$ and $\gamma_-$ are comparable, which results in the increasing contributions of the $A_+$-jumps. This is true even at larger $g$.

\section{Conclusion.}
\label{conclusion}
In this work, we have developed the CF method to calculate the pdfs of the inclusive work for the adiabatic QMMEs. We have shown that this method is also useful in discussing the symmetry of the pdfs. Hence, the CF method provides us with an alternative way of studying the quantum work besides the simulation of the quantum jump. The quantum master equations and the quantum-jump theory were known to be complementary either in the physical interpretations or in the practical calculations. Our efforts here and before may be thought of as concrete manifestations of this spirit about the theme of quantum work. Although the adiabatic quantum master equation and the equations that are about the systems driven by weak fields are two often used time-dependent QMMEs, {\it e.g.}, in the intriguing quantum heat engines~\cite{Kosloff2013}, they do not definitely cover all situations. For instance, other master equations have been proposed for the dissipative systems under intensive and fast varying fields~\cite{Cohen-Tannoudji1977,Geva1995,Kohler1997,Breuer1997,Szczygielski2013}. It shall be interesting to investigate in future whether a quantum work can be physically defined and what methods can effectively calculate it.   \\

{\noindent  We appreciate Dr. Amin for his helpful discussions about the adiabatic QMMEs. We are also grateful to Dr. Horowitz and Zhiyu Lu for reading the manuscript. This work was supported by the National Science Foundation of China under Grant No. 11174025.} 

\section*{Appendix II: Derivations of Eqs.~(\ref{workexpansion1stmoment}) and~( \ref{workexpansion2ndmoment})}
\label{appendixI}
For the first equation we write down the explicit expression of its left-hand side, 
\begin{eqnarray}
\label{explicitavgwork}
E[W]=E\left[\varepsilon_\delta(t_f)\right]-
E\left[\varepsilon_\alpha(0)\right]-\int_0^{t_f}\hbar\omega_{t_1}E\left[dN_+(t_1)\right]
+\int_0^{t_f}\hbar\omega_{t_1}E\left[dN_-(t_1)\right].
\end{eqnarray}
On the basis of the following relations~\cite{Breuer2002}, 
\begin{eqnarray}
E\left[\varepsilon_\delta(t)\right]&=&{\rm Tr}\left[H(t)\rho(t)\right],\\
E\left[dN_\pm(t)\right]&=&\gamma_\pm(\omega_{t}){\rm
Tr}\left[A_{\mp} (t)A_{\pm}(t)\rho(t)\right]dt,
\end{eqnarray}
Eq.~(\ref{explicitavgwork}) can be rewritten as
\begin{eqnarray}
E\left[W\right]&=&{\rm Tr}\left[H(t_f)\rho(t_f)\right]-{\rm Tr}\left[H(0)\rho(0)\right]+\int_0^{t_f}d\tau\hbar\omega_\tau{\rm Tr}\left[\left(\gamma_-(\omega_{t_1})A_+(t_1)A_-(t_1)-\gamma_+(\omega_{t_1})A_-(t_1)A_+(t_1)\right)\rho(t_1)\right]\nonumber\\
&=&\int_0^{t_f}dt_1\frac{d}{dt_1}{\rm Tr}\left[H(t_1)\rho(t_1)\right]-\int_0^{t_f}dt_1{\rm Tr}\left[D^\star_{t_1}[H(t_1)]\rho(t_1)\right]\nonumber\\
&=&\int_0^{t_f}dt_1{\rm Tr}\left[\partial_{t_1} H(t_1)\rho(t_1)\right],
\end{eqnarray}
where $D^\star_{t_1}$ is the adjoint superoperator of $D_{t_1}$ in Eq.~(\ref{dissipationterm}). We see that the last two equations are just the {\it first law} of thermodynamics for the adiabatic quantum master equation~(\ref{QAMME})~\cite{Alicki1979}. The derivation of  Eq.~(\ref{workexpansion2ndmoment}) is more complicated. We first write the explicit form of the left-right hand of the equation
\begin{eqnarray}
\label{secondmomentexplicitform}
E[W^2]&=&E\left [\left (\varepsilon_\delta(t_f)-\varepsilon_\alpha(0)\right)^2\right] +E\left [\left (\int_0^{t_f}dt_1\hbar\omega_{t_1}dN_-(t_1)-\int_0^{t_f}dt_2\hbar\omega_{t_2}dN_+(t_2) \right)^2\right] \nonumber\\
&&+2E\left[\left (\varepsilon_\delta(t_f)-\varepsilon_\alpha(0)\right )\left (\int_0^{t_f}dt_1\hbar\omega_{t_1}dN_-(t_1)-\int_0^{t_f}dt_2\hbar\omega_{t_2}dN_+(t_2) \right )\right].
\end{eqnarray}
In order to express these terms into the multiple time correlation functions of the operators, which is now two time points, we need exploit the following relations:
\begin{eqnarray}
E\left[\varepsilon^2_\delta(t)\right]&=&{\rm Tr}\left[H^2(t)\rho(t)\right],\\
E\left[\varepsilon_\delta(t)\varepsilon_\alpha(0)\right]&=&{\rm Tr}\left[H(t)G(t,0)H(0)\rho(0)\right],\\
E\left[\varepsilon_\delta(t_f)dN_\pm(t)\right]&=&\gamma_{\pm}(\omega_{t}){\rm Tr}\left[H(t_f)G(t_f,t)A_\pm(t)\rho(t)A_\mp(t)\right]dt,\\
E\left[\varepsilon_\alpha(0)dN_\pm(t)\right]&=&\gamma_{\pm}(\omega_{t}){\rm Tr}\left[A_\mp(t)A_\pm(t)G(t,0)H(0)\rho(0)\right]dt.
\end{eqnarray}
The other three correlation functions about $E[dN_{\pm}(t_1)dN_{\pm}(t_2)]$ have been given in our previous study~\cite{Liu2014}. Substituting them in Eq.~(\ref{secondmomentexplicitform}) and doing a careful algebra, we may arrive at the right-hand side of Eq.~(\ref{workexpansion2ndmoment}). Some details are almost parallel with what we did in the case of quantum BKE~\cite{Liu2014}. 

\section*{Appendix II: Proof of 
Eq.~(\ref{timereversalpropagators})}
\label{appendixII}
This relation between the propagator $\widetilde{{\breve{G}}}(s,0;\mu)$ and the adjoint propagator ${\breve G}^\star(t,t_f;-\nu)$ is a consequence of the characteristics of their generators,  
\begin{eqnarray}
\label{timereversalofgenerators}
\widetilde{\breve{\cal L}}_s(\mu)({O})=\Theta {\breve{\cal L}}_t^\star(-\nu)\left[\Theta^\dag{O}\Theta\right]\Theta^\dag.
\end{eqnarray}
$\widetilde{\breve{\cal L}}_s(\mu)$ has been given in   Eq.~(\ref{CFnewevolutioneq}). We need to write out another only,  
\begin{eqnarray}
\breve{\cal L}_t^\star(\mu)O=\frac{i}{\hbar}\left[H(t),O\right] &+&\sum_{\alpha=\pm}\gamma_\alpha(\omega_t)\left[e^{\alpha i \mu\hbar{\omega}_t}A_{\alpha}(t){O} A^\dag_{\alpha}(t)-\frac{1}{2}\left\{A^\dag_{\alpha}(t)A_{\alpha}(t),O\right\}\right]\nonumber\\
&+&\gamma_0\left[A_0(t){O} A^\dag_0(t)-\frac{1}{2}\left\{A^\dag_0(t)A_0(t),O\right\}\right].
\end{eqnarray} 
Then the verification of Eq.~(\ref{timereversalofgenerators}) is straightforward. Note that $\breve{\cal L}_t^\star(\mu)$ recovers the adjoint superoperator ${\cal L}_t^\star$ in Eq.~(\ref{adjointoperator}) if $\mu$ is replaced by $-i\beta$.

\bibliography{RFsubmission}

\end{document}